\documentstyle[preprint,aps,epsfig]{revtex}
\newcommand{\ra}{\rangle}
\newcommand{\la}{\langle}

\tightenlines

\begin{document}

\title{Electromagnetic couplings of the chiral perturbation theory 
Lagrangian from the perturbative chiral quark model} 

\author{V. \ E. \ Lyubovitskij \footnotemark[1],  
Th. \ Gutsche \footnotemark[1], Amand Faessler \footnotemark[1]
and R. \ Vinh Mau \footnotemark[2]  
\vspace*{0.4\baselineskip}}
\address{
\footnotemark[1] 
Institut f\"ur Theoretische Physik, Universit\"at T\"ubingen, \\
Auf der Morgenstelle 14,  D-72076 T\"ubingen, Germany 
\vspace*{0.2\baselineskip}\\
\footnotemark[2]
Laboratoire de Physique Nucleaire et des Hautes Energies, \\ 
Universit\'e P. et M. Curie, 4 Place Jussieu, 75252 Paris Cedex 05, France 
\vspace*{0.3\baselineskip}\\}

\maketitle 

\vskip.5cm

\begin{abstract}
We apply the perturbative chiral quark model to the study of the low-energy 
$\pi N$ interaction. Using an effective chiral Lagrangian we reproduce the 
Wein\-berg-Tomozawa result for the $S$-wave $\pi N$ scattering lengths. After 
inclusion of the photon field we give predictions for the electromagnetic 
$O(p^2)$ low-energy couplings of the chiral perturbation theory effective 
Lagrangian that define the electromagnetic mass shifts of nucleons and 
first-order $(e^2)$ radiative corrections to the $\pi N$ scattering amplitude. 
Finally, we estimate the leading isospin-breaking correction to the strong 
energy shift of the $\pi^- p$ atom in the $1s$ state, which is relevant for 
the experiment "Pionic Hydrogen" at PSI.   
\end{abstract}

\vspace*{\baselineskip}
\vskip1cm

       

\section{Introduction} 

Hadron models set up to understand the structure of the nucleon should 
respect the constraints imposed by chiral symmetry. Spontaneous and explicit 
chiral symmetry breaking requires the existence of the pion whose mass 
vanishes in the limit of zero current quark mass. In turn, nucleon 
observables have to receive contributions from the pion cloud. 

We recently suggested \cite{PCQM} a baryon model as based on earlier 
consideration \cite{Gutsche}, the perturbative chiral quark model (PCQM), 
which includes relativistic quark wave functions and confinement as well as 
the chiral symmetry requirements. The PCQM was successfully applied to 
$\sigma$-term physics \cite{PCQM} and extended to the study of electromagnetic 
properties of the nucleon. Similar perturbative quark models have also been 
studied in \cite{Thomas-Chin}. Although the Lagrangian for this model fulfils 
the $SU(2) \times SU(2)$ symmetry, it is in general nontrivial to identify the 
explicit hadron dynamics which leads to well-known predictions of current 
algebra for various observables. For instance, the $S$-wave $\pi N$ scattering 
at threshold is such a process in the presence of a pion cloud. 

As was shown by Weinberg and Tomozawa \cite{Weinberg1,Tomozawa}, 
a model-independent expression for the $S$-wave $\pi N$ scattering lengths 
in terms of the pion mass and the weak decay constant is obtained when using 
the current algebra relations and the partial conservation of axial current 
(PCAC) assumption:   
\begin{eqnarray}\label{scatt_lengths} 
a_{\frac{1}{2}} = \frac{M_\pi}{4\pi F_\pi^2} + o(M_\pi)  
\,\,\,\,\, \mbox{and} \,\,\,\,\,  
a_{\frac{3}{2}} = -\frac{M_\pi}{8\pi F_\pi^2} + o(M_\pi)  
\end{eqnarray}
To reproduce this result for the $\pi N$ scattering lengths one can also use 
a specific Lagrangian in the nucleon field refered to as the Weinberg-Tomozawa 
(WT) term \cite{Weinberg2}-\cite{Leutwyler1}. Formulated on the quark 
level \cite{Thomas1,Jennings} the WT term is part of the effective Weinberg 
Lagrangian \cite{Weinberg2}. This form can be derived from the original 
$\sigma$-model \cite{Gell-Mann_Levy} by performing a chiral-field dependent 
rotation on the quark field. Thereby, the transformation eliminates the 
nonderivative coupling of the chiral (pion) field with the quarks and replaces 
it by a nonlinear derivative coupling (axial vector term + WT term + higher 
order terms in the chiral field). Both realizations of chirally-symmetric 
Lagrangians (the original $\sigma$-model and the Weinberg type Lagrangian) 
should \'{a} priori give the same result for the $\pi N$ $S$-wave scattering 
lengths.  

Starting point of present considerations is the PCQM \cite{PCQM} which, as 
dictated by global chiral symmetry, originally employs a nonderivative 
coupling between the pions and the quarks \cite{Gell-Mann_Levy}. The purpose 
of the present work is first to demonstrate how, in the context of this chiral 
quark model, the WT result can be consistently obtained to the order of 
accuracy we are working in. Second, after fulfilling the constraint of 
reproducing $\pi N$ $S$-wave scattering at threshold, we include the photon 
field in our formalism. We determine first-order ($e^2$) radiative corrections 
to the nucleon mass and the pion-nucleon amplitude at threshold. We thereby 
predict the full set of $O(p^2)$ electromagnetic low-energy couplings (LECs) 
originally defined in the effective Lagrangian of chiral perturbation theory 
(ChPT). Quantitative information about these constants is important for the 
ongoing experimental and theoretical analysis of decay properties of the 
$\pi^-p$ atom (for a detailed discussion see Ref. \cite{piP-atom}). 
In particular, we give a prediction for the leading isospin-breaking 
correction to the strong energy shift of the $\pi^- p$ atom in the $1s$ state.

\section{Perturbative Chiral Quark Model and $\pi N$ $S$-wave scattering} 

The PCQM is based on the effective, chirally invariant Lagrangian 
${\cal L}_{inv}$ \cite{PCQM}  
\begin{eqnarray}\label{Linv} 
{\cal L}_{inv}(x)&=&\bar\psi(x) \biggl\{ i\not\! \partial -\gamma^0 V(r)
-  \, S(r) \, \biggl[ \frac{U \, + U^\dagger}{2} \, 
+ \, \gamma^5 \, \frac{U \, -  U^\dagger}{2} \biggr] \biggr\} \, \psi(x)\\
&+&\frac{F^2}{4} \, {\rm Tr} [\, \partial_\mu U \, 
\partial^\mu U^\dagger \, ]\nonumber  
\end{eqnarray}
where $\psi$ is the quark field, $U$ is the chiral field and $F=88$ MeV is the 
pion decay constant in the chiral limit \cite{PCQM,Gasser1}. The quarks move 
in a self-consistent field, represented by scalar $S(r)$ and vector $V(r)$ 
components of a static potential with $r=|\vec{x}|$ providing 
confinement. The interaction of quarks with Goldstone bosons is introduced via 
the chiral field $U=\exp[i\hat\Phi/F]$ which represents the {\it exponentional 
parametrization} of the nonlinear $\sigma$-model \cite{Gell-Mann_Levy}. 
$\hat\Phi$ is the matrix of pseudoscalar mesons (in the following we restrict 
ourselves to the $SU(2)$ flavor case, 
$\hat\Phi \to \hat\pi  = \vec{\pi} \vec{\tau}$.). For small fluctuations of 
the meson fields, one can use the expansion in powers of the parameter $1/F$. 

Usually, the Lagrangian (\ref{Linv}) is linearized with respect to the 
field $\hat\Phi$:  
\begin{eqnarray}\label{Linear}
U=\exp\biggl[i \, \frac{\hat\Phi}{F} \biggr] \simeq 
1 + i \frac{\hat\Phi}{F} + o\biggl(\frac{\hat\Phi}{F}\biggr).
\end{eqnarray}
Such an approximation is meaningful if we consider processes without external 
pions field (e.g. nucleon mass shift due to the pion cloud) or with a single 
external pion field (e.g. pion-nucleon form factor). The resulting approximate 
chiral invariance of the linearized Lagrangian \cite{PCQM,Thomas1} guarantees 
both a conserved axial current (or PCAC in the presence of the meson mass 
term) and the Goldberger-Treiman relation \cite{PCQM,Thomas1}. 

However, to calculate the pion-nucleon scattering amplitude, for consistency, 
we have to expand the invariant Lagrangian (\ref{Linv}) up to quadratic terms 
in the pion field. Adding the mass term for pions we obtain the effective 
Lagrangian ${\cal L}_{eff}$:   
\begin{eqnarray}\label{Linv_expand}
{\cal L}_{eff}(x)&=& {\cal L}_0(x) \, + \, {\cal L}^{str}_I(x) 
\, + \, o(\vec{\pi}^2) ,\\ 
{\cal L}_0(x) &=& \bar\psi(x) \biggl\{ i\not\! \partial - S(r) 
- \gamma^0 V(r) \biggr\} \psi(x) 
- \frac{1}{2} \, \vec{\pi}(x) \, (\Box + M_\pi^2) \,\vec{\pi}(x) , \nonumber\\ 
{\cal L}^{str}_I(x)&=& - \frac{S(r)}{F} \bar\psi(x) \, i \, \gamma^5 \, 
\vec{\tau} \, \vec{\pi}(x) \, \psi(x) + \frac{S(r)}{2F^2} \bar\psi(x) \, 
\vec\pi^2(x)  \, \psi(x) . \nonumber  
\end{eqnarray}
where $\Box = \partial^\mu \partial_\mu$ and $M_\pi$ is the pion mass.  

The quark field $\psi$ we expand in the basis of potential eigenstates as 
\begin{eqnarray}\label{total_psi}
\psi(x) &=& \sum\limits_\alpha b_\alpha u_\alpha(x)
+ \sum\limits_\beta d_\beta^\dagger v_\beta(x) , \\
u_\alpha(x) &=& u_\alpha(\vec{x}) \exp(-i{\cal E}_\alpha t), \hspace*{.5cm}
v_\beta(x) = v_\beta(\vec{x}) \exp(i{\cal E}_\beta t) . \nonumber
\end{eqnarray}
The sets of quark $\{ u_\alpha \}$ and antiquark $\{ v_\beta \}$ wave 
functions in orbits $\alpha$ and $\beta$ are solutions of the static 
Dirac equation with 
\begin{eqnarray}\label{Dirac_Eq}
& &\hspace*{1cm}[-i\vec{\gamma}\vec{\nabla}+S(r)+
\gamma^0(V(r)-{\cal E}_\alpha)]u_\alpha(\vec{x})=0  \\
\mbox{and}& &\hspace*{1cm}
[-i\vec{\gamma}\vec{\nabla}+S(r)+\gamma^0(V(r)+{\cal E}_\beta)] 
v_\beta(\vec{x})=0 . \nonumber
\end{eqnarray}    
The wave functions satisfy to the normalization
\begin{eqnarray}\label{normalization}
\int d^3x \, u^{\dagger}_{\alpha^\prime}(\vec{x}) u_\alpha(\vec{x}) 
= \delta_{\alpha\alpha^\prime}, 
\hspace*{1cm} 
\int d^3x \, v^{\dagger}_{\beta^\prime}(\vec{x}) v_\beta(\vec{x}) 
= \delta_{\beta\beta^\prime}, 
\end{eqnarray}
and completeness conditions 
\begin{eqnarray}\label{completeness}
\sum\limits_{\alpha} u_\alpha(\vec{x}) u_\alpha^\dagger(\vec{y}) + 
\sum\limits_{\beta} v_\beta(\vec{x}) \bar v_\beta^\dagger(\vec{y}) = 
\delta^{(3)}(\vec{x}-\vec{y}) .  
\end{eqnarray}
The expansion coefficients $b_\alpha$ and $d_\beta^\dagger$ are the 
corresponding single quark annihilation and antiquark creation operators, 
which fulfil the usual canonical anticommutation relations 
\begin{eqnarray}\label{canon_anti}
\{ b_\alpha, b^\dagger_{\alpha^\prime} \} = \delta_{\alpha\alpha^\prime} 
\hspace{.5cm} \mbox{and} \hspace{.5cm}
\{ d_\beta,  d^\dagger_{\beta^\prime}  \}    = \delta_{\beta\beta^\prime} . 
\end{eqnarray}
The quark propagator in the binding potential is:  
\begin{eqnarray}\label{quark_propagator}  
i G_\psi(x,y) &=& <0|T\{\psi(x)\bar\psi(y)\}|0> \\
&=&\sum\limits_{\alpha} u_\alpha(x) \bar u_\alpha(y) \theta(x_0-y_0)  
- \sum\limits_{\beta} v_\beta(x) \bar v_\beta(y) \theta(y_0-x_0) .\nonumber  
\end{eqnarray}

In the PCQM the strong pion-nucleon amplitude 
is defined as\footnote{Here and in the following we use the interaction 
Lagrangian and Wick's $T$-ordering \cite{Bogoliubov_Shirkov} in the 
calculation of matrix elements.}   
\begin{eqnarray}\label{piN_strong}
{^N\la}\phi_0; \pi_j| \, \sum\limits_{n=1}^{2} \frac{i^n}{n!} 
\int  \, d^4x_1 \ldots \int  d^4x_n \, \, 
T[{\cal L}_{I}^{str}(x_1) \ldots {\cal L}_I^{str}(x_n) \, ] \, 
|\phi_0; \pi_i{\ra^N_c} 
\end{eqnarray}
where the state vector $|\phi_0; \pi_i\ra$ corresponds to the three quark 
ground state $\phi_0$ and a pion. Subscript "$c$" in Eq. (\ref{piN_strong}) 
refers to contributions from connected graphs only. Superscript "$N$" 
indicates that the matrix elements have to be projected onto the respective 
nucleon states. These nucleon states are conventionally set up by the product 
of single quark $SU(6)$ spin-flavor and $SU(3)_c$ color wave functions (see 
details in \cite{Close}), where the nonrelativistic single quark spin wave 
function is replaced by the relativistic ground state solution. At the tree 
level approximation (when we neglect pion loops) three diagrams (Fig.1) 
contribute to the $\pi N$ scattering amplitude: the $s$-channel pole (Fig.1a), 
the $u$-channel pole (Fig.1b) and the seagull diagram (Fig.1c). Pole diagrams 
are generated by the $\pi$-quark pseudoscalar coupling, whereas the seagull 
diagram is due to the quadratic term in the pion field. 

To evaluate the matrix element (\ref{piN_strong}) we use a set of identities 
deduced from the Dirac equation (\ref{Dirac_Eq}) and the completeness 
condition (\ref{completeness}): 
\begin{eqnarray}\label{Dirac_eqs0} 
\bar\psi(x) i \gamma^5 S(|\vec{x}|) \psi(x) = \frac{1}{2} \,\, \partial_\mu^x 
\{ \bar\psi(x) \gamma^\mu \gamma^5 \psi(x) \}  
\end{eqnarray}
and
\begin{eqnarray}\label{Dirac_eqs} 
\bar\psi(x) i \gamma^5 S(|\vec{x}|) i G_\psi(x,y) = 
\frac{1}{2} \,\, \partial_\mu^x 
\{ \bar\psi(x) \gamma^\mu \gamma^5 i G_\psi(x,y) \} 
+ \frac{1}{2} \bar\psi(x) \gamma^5 \delta^{(4)}(x-y) . 
\end{eqnarray}
Using Eq.~(\ref{Dirac_eqs}) we obtain the key identity for the evaluation 
of Eq.~(\ref{piN_strong}):  
\begin{eqnarray}\label{Key_piN_thr}
\hspace*{-.2cm}
& &4 \, \bar\psi(x) i\gamma^5 S(|\vec{x}|) \tau_i i G_\psi(x,y) 
i\gamma^5 S(|\vec{y}|) \tau_j \psi(y) 
= 2 \, i \, \delta^{(4)}(x-y) 
\bar\psi(x) \tau_i \tau_j S(|\vec{x}|) \psi(x)  \\[.2cm]
\hspace*{-.2cm}
&+& \partial_\mu^x [ \delta^{(4)}(x-y) 
\bar\psi(x) \gamma^\mu \tau_i \tau_j \psi(x) ] 
+ \partial_\mu^x  \partial_\nu^y 
[ \bar \psi(x) \gamma^\mu \gamma^5 \tau_i  
  i G_\psi(x,y) \gamma^\nu \gamma^5 \tau_j \psi(y) ] . \nonumber 
\end{eqnarray}
With help of Eq.~(\ref{Key_piN_thr}) the contributions to the $\pi N$ 
amplitude from the pole (PL) diagrams (Fig.1a and Fig.1b) are given by:  
\begin{eqnarray}\label{T_pole} 
& &2\pi i \, \delta(p_1^0-p_2^0) \, \chi^\dagger_{N_{f^\prime}}   \, 
T^{ij; {\rm PL}}_{\pi N} \, \chi_{N_f} =  
{^N\la}\phi_0; \pi_j| \sum\limits_{I=1}^{3} \hat O_I |\phi_0; \pi_i{\ra^N} \\ 
& &\hat O_1 = - \frac{i}{4F^2} \, \int d^4x \, \bar\psi(x) \gamma^\mu 
[\vec{\pi}(x) \times \partial_\mu \vec{\pi}(x)] \vec{\tau} \psi(x) \nonumber\\ 
& &\hat O_2 = - \frac{1}{4F^2} \, \int d^4x \int d^4y \, 
\bar\psi(x) \gamma^\mu \gamma^5 \partial_\mu\hat{\pi}(x) i G_\psi(x,y) 
\gamma^\nu \gamma^5 \partial_\nu\hat{\pi}(y) \psi(y) \nonumber\\
& &\hat O_3 = - \frac{i}{2F^2} \,\, \int d^4x \, \bar\psi(x) 
S(|\vec{x}|) \vec{\pi}^{\, 2} \psi(x) \nonumber . 
\end{eqnarray}
where $\chi_{N_f}$ and $\chi^\dagger_{N_{f^\prime}}$ are the nucleon isospin 
wave functions in the initial and final states, $p_1^0$ and $p_2^0$ are the 
energies of incoming and outcoming pion, respectively. 
The seagull (SG) graph results in the expression 
\begin{eqnarray}\label{T_seagull}
2\pi i \, \delta(p_1^0-p_2^0) \, \chi^\dagger_{N_{f^\prime}}  
\, T^{ij; \rm SG}_{\pi N} \, 
\chi_{N_f} = \,\, - \,\, {^N\la}\phi_0; \pi_j| \hat O_3 |\phi_0; \pi_i{\ra^N} ,
\end{eqnarray}
which cancels the corresponding term in Eq. (\ref{T_pole}). The sum of the  
pole and seagull diagrams is then: 
\begin{eqnarray}\label{sum_piN}
\hspace*{-.5cm}
2\pi i\, \delta(p_1^0-p_2^0) \,\chi^\dagger_{N_{f^\prime}}\, T_{\pi N}^{ij} \, 
\chi_{N_f} &=& 2\pi i\, \delta(p_1^0-p_2^0) \,\chi^\dagger_{N_{f^\prime}}\,\, 
[\, T^{ij; {\rm PL}}_{\pi N} \, + \, T^{ij; {\rm SG}}_{\pi N} \,]\,\chi_{N_f}\\
\hspace*{-.5cm} 
&=& \, {^N\la}\phi_0; \pi_j| \hat O_1 + \hat O_2 |\phi_0; \pi_i{\ra^N} . 
\nonumber 
\end{eqnarray}
which is identical to the pion-nucleon amplitude generated by the Weinberg 
Lagrangian. The $\pi N$ scattering amplitude is: 
\begin{eqnarray}\label{piN_thr}
T^{ij}_{\pi N} = T^{ij; \, {\rm PL}}_{\pi N} + T^{ij; \, {\rm SG}}_{\pi N} = 
- \frac{i\varepsilon^{ijk}}{2F^2} \tau^k M_\pi\, + \, 
O\biggl( \frac{\vec{p}^{\, 2}}{M_\pi}, M_\pi^2 \biggr) ,      
\end{eqnarray}
where $O(\vec{p}^{\, 2}/M_\pi, M_\pi^2)$ represents a contribution which 
vanishes at threshold (with $\pi$ three-momentum $|\,\vec{p}\,| \to 0$) and 
faster than $M_\pi$. We, therefore, recover the result of Weinberg and 
Tomozawa for the $S$-wave $\pi N$ scattering lengths (\ref{scatt_lengths}). 
Note, that only intermediate antiquark states contribute to the $\pi N$ 
amplitude at threshold, whereas the quark contribution (as in a relativistic 
theory with a fully covariant quark propagator) is vanishing. Also, due to the 
normalization (\ref{normalization}) and the completeness condition 
(\ref{completeness}), the explicit form of the quark and antiquark 
solutions is not needed in the derivation of (\ref{piN_thr}). 

Actually, one can generate the Weinberg-Tomozawa term at the Lagrangian level 
through the unitary chiral transformation 
$\psi \to \exp\{ -i\gamma^5\hat{\Phi}/(2F) \} \psi$ on the quark field $\psi$, 
as done, for example, in the cloudy bag model \cite{Thomas1}. With this 
unitary chiral rotation the Lagrangian (\ref{Linv}) 
transforms into a Weinberg-type form ${\cal L}^W$ 
\begin{eqnarray}\label{L_W}  
\hspace*{-.75cm}{\cal L}^W(x) &=& {\cal L}_0(x) \, + \, 
{\cal L}^{W; str}_I(x) \, + \, o(\vec{\pi}^2) ,\\ 
\hspace*{-.75cm}{\cal L}_0(x) &=& \bar\psi(x) \biggl\{ i\not\! \partial - S(r) 
- \gamma^0 V(r) \biggr\} \psi(x) 
- \frac{1}{2} \, \vec{\pi}(x) \, (\Box + M_\pi^2) \,\vec{\pi}(x) , \nonumber\\ 
\hspace*{-.75cm}
{\cal L}^{W; str}_I(x) &=& \frac{1}{2F} \partial_\mu \vec{\pi}(x) \, 
\bar\psi(x) \, \gamma^\mu \, \gamma^5 \, \vec{\tau} \, \psi(x) 
- \frac{\varepsilon_{ijk}}{4F^2} \, \pi_i(x) \, \partial_\mu \pi_j(x) \, 
\bar\psi(x) \, \gamma^\mu \, \tau_k \,  \psi(x) , \nonumber
\end{eqnarray}
which contains now an axial vector coupling and the WT term. Thereby 
${\cal L}^{W; str}_I$ is the $O(\pi^2)$ strong interaction Lagrangian. 
Since the Lagrangians ${\cal L}_{inv}$ and ${\cal L}^W$ (up to order 
$1/F^2$) are connected by a unitary transformation, both have to result in 
an identical expression for the $\pi N$ scattering amplitude, when evaluated 
up to a given order in $1/F$. To make the connection explicit we start from 
the original Lagrangian (\ref{Linv}). We show that the Weinberg-Tomozawa 
result can be reproduced if: i) we use the expansion of the  chiral field up 
to quadratic terms and ii) we employ the full quark propagator including the 
antiquark components. 

Thus we have demonstrated explicitly for the $\pi N$ amplitude up to order 
$(1/F^2)$ that the two effective theories, the one involving the pseudoscalar 
coupling and the Weinberg type, are formally equivalent, both on the level of 
the Lagrangians and for the matrix elements. This equivalence is based on the 
unitary transformation of the quark fields, when, in addition, the quarks 
remain on their energy shell. The same relation also holds in a fully 
covariant formalism, when quarks/baryons are on their mass shell. Having set 
the basis in the strong interaction sector we will indicate in the next 
section that the two forms of the interaction Lagrangian also yield the same 
results when including the photon field. 

\section{Applications: electromagnetic corrections to \\
the nucleon mass and the $\pi N$ amplitude at threshold}

We can now use the equivalence between Eq. (\ref{Linv_expand}) and (\ref{L_W}) 
to introduce electromagnetic corrections in the PCQM. This is performed in 
an unambiguous way into Eq. (\ref{Linv_expand}): 
\begin{eqnarray} 
\hspace*{-.5cm}
\partial_\mu\psi \to D_\mu\psi = \partial_\mu\psi + i e Q A_\mu \psi,  
\hspace*{.5cm} 
\partial_\mu\pi_i \to D_\mu\pi_i = \partial_\mu\pi_i + 
e \varepsilon_{3ij} A_\mu \pi_j
\end{eqnarray}
where $A_\mu$ is the electromagnetic field and $Q$ is the quark charge matrix. 
Following the Gell-Mann and Low theorem \cite{Gell-Mann_Low} the 
electromagnetic mass shift $\Delta m_N^{em}$ of the nucleon with respect to 
the three-quark ground state $|\phi_0>^N$ is given by
\begin{eqnarray}\label{Energy_shift} 
\Delta m_N^{em} \doteq  \,\, {^N\la}\phi_0| \, - \frac{i}{2} \, 
\int \, \delta(x^0) \, d^4x \, \int \, d^4y \, 
T[{\cal L}^{em}(x){\cal L}^{em}(y)] \, |\phi_0{\ra^N_c} 
\end{eqnarray}
to order $e^2$ in the electromagnetic interaction. Subscript "$c$" in 
Eq. (\ref{Energy_shift}) refers to contributions from connected graphs only. 
With the quark-photon interaction defined by the Lagrangian 
\begin{eqnarray}
{\cal L}^{em}(x) = - e A_\mu \bar\psi(x) Q \gamma^\mu \psi(x) , 
\end{eqnarray} 
the electromagnetic mass shift $\Delta m_N^{em}$ is generated by two diagrams: 
one-body (Fig.2a) and two-body diagram (Fig.2b).  

The $\pi N$ scattering amplitude at threshold including the leading 
electromagnetic corrections (up to order $e^2/F^2$) is generated by the 
interaction Lagrangian 
\begin{eqnarray}\label{L_int}  
{\cal L}_I(x) \, = \, {\cal L}_I^{str}(x) \, + \, {\cal L}_I^{em}(x) .  
\end{eqnarray}
The strong interaction part ${\cal L}_I^{str}$ is already defined in 
Eq. (\ref{Linv_expand}) and ${\cal L}_I^{em}$ is the $O(e)$ electromagnetic 
Lagrangian describing the interaction of quarks and pions with the 
electromagnetic field 
\begin{eqnarray}\label{L_int_em}
{\cal L}_I^{em} \, = \, - e A_\mu(x) \bar\psi(x) Q \gamma^\mu \psi(x) \, 
- \, e \varepsilon_{3ij} A_\mu(x) \pi_i(x) \partial^\mu \pi_j(x) . 
\end{eqnarray} 
The $\pi N$ amplitude in the presence of $O(e^2)$ radiative corrections 
is given by 
\begin{eqnarray}\label{Key_Eq}
{^N\la}\phi_0; \pi_j| \, \sum\limits_{n=1}^{4} 
\frac{i^n}{n!} \int  \, d^4x_1 \ldots \int  d^4x_n \, \, 
T[{\cal L}_{I}(x_1) \ldots {\cal L}_I(x_n) \, ] \, |\phi_0; \pi_i{\ra^N_c} . 
\end{eqnarray}
Using Eq. (\ref{Dirac_eqs}) we obtain the identities
\begin{eqnarray}\label{Key_Eq_new1}
& &2 \, i G_\psi(x,y) i\gamma^5 S(|\vec{y}|) \tau_i i G_\psi(y,z)\\[.2cm] 
\hspace*{-.2cm}
&=& \delta^{(4)}(x-y) \gamma^5 \tau_i i G_\psi(y,z) 
 +  \delta^{(4)}(y-z) i G_\psi(x,y) \gamma^5 \tau_i 
 +  \partial_\mu^y \Gamma^\mu_i(x,y,z) 
\nonumber 
\end{eqnarray}
and 
\begin{eqnarray}\label{Key_Eq_new2}
& &4 \, i G_\psi(x,y) i\gamma^5 S(|\vec{y}|) \tau_i i G_\psi(y,z) 
i\gamma^5 S(|\vec{z}|) \tau_j i G_\psi(z,w) \\[.2cm]
\hspace*{-.2cm}
&=&  2 \, i \delta^{(4)}(y-z) i G_\psi(x,y) \tau_i \tau_j 
    S(|\vec{y}|) i G_\psi(z,w) +  
    \delta^{(4)}(x-y)\delta^{(4)}(z-w) \gamma^5 \tau_i 
    i G_\psi(x,w ) \gamma^5 \tau_j \nonumber\\ [.2cm]
&+& \delta^{(4)}(x-y)\delta^{(4)}(y-z) \tau_i\tau_j  i G_\psi(x,w)
 +  \delta^{(4)}(y-z)\delta^{(4)}(z-w) i G_\psi(x,w)\tau_i\tau_j
 \nonumber\\[.2cm]
 \hspace*{-.2cm}
&+& \delta^{(4)}(z-w) \partial_\mu^y\Gamma^\mu_i(x,y,z) \gamma^5 \tau_j 
 + \delta^{(4)}(x-y)\gamma^5 \tau_i \partial_\mu^z\Gamma^\mu_j(x,z,w) 
 + \partial_\mu^y \partial_\nu^z \Gamma^{\mu\nu}_{ij}(x,y,z,w)\nonumber 
\end{eqnarray} 
where 
\begin{eqnarray}
\Gamma^\mu_i(x,y,z) 
&=& i G_\psi(x,y) \gamma^\mu\gamma^5 \tau_i i G_\psi(y,z)\\
\Gamma^{\mu\nu}_{ij}(x,y,z,w) &=& i G_\psi(x,y) \gamma^\mu\gamma^5 \tau_i 
i G_\psi(y,z) \gamma^\nu\gamma^5 \tau_j i G_\psi(z,w) . \nonumber
\end{eqnarray}
With help of Eqs. (\ref{Key_Eq_new1}) and (\ref{Key_Eq_new2}) one can 
show that the matrix element (\ref{Key_Eq}) is identical to the one where 
${\cal L}_I$ is replaced by the Weinberg interaction Lagrangian including 
photons ${\cal L}^W_I$ 
\begin{eqnarray}\label{L_int_W}  
{\cal L}_I^{W}(x) = {\cal L}_I^{W; str}(x) + {\cal L}_I^{W; em}(x)
\end{eqnarray} 
where ${\cal L}_I^{W; str}$ is given in Eq. (\ref{L_W}) and the additional 
electromagnetic part ${\cal L}_I^{W; em}$ is given by 
\begin{eqnarray}
{\cal L}_I^{W; em}(x) &=& {\cal L}_I^{em}(x) \, + \, 
\frac{e \varepsilon_{3ij}}{2F} A_\mu(x) \pi_j(x) 
\bar\psi(x) \gamma^\mu \gamma^5 \tau_i \psi(x) \\   
\hspace*{-1cm}
&+&  \frac{e}{4F^2} A_\mu(x) \bar \psi(x) \gamma^\mu 
[ \vec{\pi}^{\, 2}(x) \tau_3 - \vec{\pi}(x) \vec{\tau} \pi^0(x) ] \psi(x) . 
\nonumber
\end{eqnarray} 
We therefore have the key identity 
\begin{eqnarray}\label{Matrix_Key_Eq} 
& &{^N\la}\phi_0; \pi_j| \, \sum\limits_{n=1}^{4} \frac{i^n}{n!} 
\int  \, d^4x_1 \ldots \int  d^4x_n \, \, 
T[{\cal L}_{I}(x_1) \ldots {\cal L}_I(x_n) \, ] \, |\phi_0; \pi_i{\ra^N_c} \\
&=&{^N\la}\phi_0; \pi_j| \, \sum\limits_{n=1}^{4} \frac{i^n}{n!} 
\int  \, d^4x_1 \ldots \int  d^4x_n \, \, 
T[{\cal L}_{I}^{W}(x_1) \ldots {\cal L}_I^{W}(x_n) \, ] \, 
|\phi_0; \pi_i{\ra^N_c} \nonumber 
\end{eqnarray}
extending the consistency between the two interaction Lagrangians to the case 
where photons are present. For the equivalence to hold is essential that the 
photons are introduced consistently in both formalisms, that is by minimal 
substitution. The diagrams for $O(e^2/F^2)$ radiative corrections to the 
$\pi N$ amplitude at threshold are shown in Fig.3. They are generated by the 
matrix element (\ref{Matrix_Key_Eq}) with the Weinberg interaction Lagrangian 
${\cal L}^W_I$. Again, the set of diagrams of Fig.3 can be identically deduced 
from the corresponding one generated by our original interaction Lagrangian 
${\cal L}_I$. To evaluate the diagrams shown in Figs.2 and 3 we use the photon 
propagator in the Coulomb gauge\footnote{It can be shown that the results do 
not depend on the choice of the gauge.} to separate the contributions from 
Coulomb ($A_0$) and transverse ($A_i$) photons. The propagators of Coulomb 
($D_{00}$) and transverse ($D_{ij}$) photons are given by 
\begin{eqnarray}
\hspace*{-.75cm}
iD_{00}(x-y)&=&<0|T\{A_0(x)A_0(y)\}|0> = i \int\frac{d^4k}{(2\pi)^4}
\frac{e^{-ik(x-y)}}{\vec{k}^{\, 2}}  
\end{eqnarray}
and 
\begin{eqnarray}
\hspace*{-.75cm}
iD_{ij}(x-y)&=&<0|T\{A_i(x)A_j(y)\}|0> = 
i \int\frac{d^4k}{(2\pi)^4} \frac{e^{-ik(x-y)}}{k^2 + i \varepsilon} 
\biggl\{ \delta_{ij} - \frac{k_ik_j}{\vec{k}^{\, 2}} \biggr\}. 
\end{eqnarray}
First, we analyze the electromagnetic mass shift of the nucleon 
$(\Delta m^{em}_N)$. The contributions of diagrams Fig.2a (one-body term 
$\Delta m^{em; a}_N$) and Fig.2b (two-body term $\Delta m^{em; b}_N$) are 
given by  
\begin{eqnarray}\label{one_full}
\Delta m^{em; a}_N = e^2 \, \cdot \, {^N\la}\phi_0| \int d^4x \int d^4y \, 
\delta(x^0) D_{\mu\nu}(x-y) \bar \psi_0(x) \gamma^\mu Q iG_\psi(x,y) 
\gamma^\nu Q \psi_0(y) |\phi_0{\ra^N} 
\end{eqnarray}
and 
\begin{eqnarray}\label{two_full}
\Delta m^{em; b}_N &=& \frac{e^2}{2} \, \cdot \, {^N\la}\phi_0| \int d^4x 
\int d^4y \, \delta(x^0) D_{\mu\nu}(x-y) \bar \psi_0(x) \gamma^\mu Q 
\psi_0(x) \bar \psi_0(y) \gamma^\nu Q \psi_0(y) |\phi_0{\ra^N} . 
\end{eqnarray}
In the following we truncate the expansion of the quark propagator to the 
ground state eigen mode:  
\begin{eqnarray}\label{quark_propagator_ground}  
iG_\psi(x,y) \to iG_0(x,y) \doteq u_0(\vec{x}) \bar u_0(\vec{y}) 
e^{-i{\cal E}_\alpha(x_0-y_0)}\theta(x_0-y_0), 
\end{eqnarray}
that is we restrict the intermediate baryon states to $N$ and $\Delta$ 
configurations. Inclusion of excited baryon states will be subject of future 
investigations. With the approximation (\ref{quark_propagator_ground}),  
the one- and two-body contributions reduce to 
\begin{eqnarray}\label{one} 
\Delta m^{em; a}_N = \frac{e^2}{16\pi^3} \la N| \sum\limits_{i=1}^3 
(Q^2)^{(i)} |N\ra \int \frac{d^3q}{\vec{q}^{\,\, 2}} 
\biggl\{ [G_E^p(-\vec{q}^{\, 2})]^2 - \frac{\vec{q}^{\,\, 2}}{2m_N^2} 
[G_M^p(-\vec{q}^{\, 2})]^2 \biggr\} 
\end{eqnarray}
and 
\begin{eqnarray}\label{two}
\Delta m^{em; b}_N &=& \frac{e^2}{16\pi^3} \int 
\frac{d^3 q}{\vec{q}^{\,\, 2}} \, \, 
\biggl\{ \la N|\sum\limits_{i\,\not\! = j}^3 
Q^{(i)}Q^{(j)}|N\ra [G_E^p(-\vec{q}^{\, 2})]^2 \\ 
&-&\la N|\sum\limits_{i\,\not\! = j}^3 Q^{(i)} Q^{(j)} \vec{\sigma}^{(i)} 
\vec{\sigma}^{(j)}|N\ra \, \frac{\vec{q}^{\,\, 2}}{6m_N^2} 
[G_M^p(-\vec{q}^{\, 2})]^2 \biggr\} , \nonumber
\end{eqnarray}
where $|N \ra$ is the SU(6) spin-flavor w.f. of the nucleon. Here we introduce 
the proton charge $(G_E^p)$ and magnetic $(G_M^p)$ form factors calculated 
at zeroth order \cite{PCQM} (meson cloud corrections are not taken into 
account) with   
\begin{eqnarray}
\chi^\dagger_{N_{s^\prime}} \chi_{N_s} G_E(-\vec{q}^{\, 2}) = 
{^N\la}\phi_0|\int d^3x \bar \psi_0(\vec{x}) \gamma^0 \psi_0(\vec{x})
e^{ i \vec{q} \, \vec{x}}|\phi_0{\ra^N} 
\end{eqnarray} 
and 
\begin{eqnarray}
\chi^\dagger_{N_{s^\prime}}
\frac{i \, [ \vec{\sigma}_N \times \vec{q} \, ]}{2m_N} 
\chi_{N_s} G_M(-\vec{q}^{\, 2}) =
{^N\la}\phi_0|\int d^3x \bar \psi_0(\vec{x}) \vec{\gamma} \psi_0(\vec{x})
e^{i \vec{q} \, \vec{x}}|\phi_0{\ra^N} 
\end{eqnarray} 
where $\chi_{N_s}$ and $\chi^\dagger_{N_{s^\prime}}$ are the nucleon spin w.f. 
in the initial and final state; $\vec{\sigma}_N$ is the nucleon spin operator. 
Note that the contributions of Coulomb and transverse photons to the 
electromagnetic mass shifts (see Eqs. (\ref{one}) and (\ref{two})) are related 
to the nucleon charge and magnetic form factors, respectively. The sum of 
expectation values  
\begin{eqnarray}
\la N|\sum\limits_{i=1}^3 (Q^2)^{(i)}|\ra N  + 
\la N|\sum\limits_{i \, \not\! = j}^3 Q^{(i)}Q^{(j)}|N \ra = \left \{
\begin{array}{cc}
1   & \mbox{for} \,\,\, N=p\\
0   & \mbox{for} \,\,\, N=n\\
\end{array}
\right. 
\end{eqnarray}
is equivalent to the charge matrix of nucleons ($Q_N$ being the nucleon 
charge). In the limit $m_N \to \infty$ (when we neglect the contribution of 
$G_M^p$ in Eqs. (\ref{one}) and (\ref{two})) we obtain for the 
electromagnetic mass shifts 
\begin{eqnarray}\label{Delta_mn}
\Delta m^{em}_N = \Delta m_N^{em; a} +  \Delta m_N^{em; b} =  
\frac{\alpha Q_N^2}{4\pi^2} 
\int \frac{d^3q}{\vec{q}^{\,\, 2}} \, [G_E^p(-\vec{q}^{\, 2})]^2  
\end{eqnarray}
consistent with the result (Eq. (12.4)) of Ref. \cite{Gasser_Leutwyler_PR}. 
Hence, the electromagnetic mass shift of the neutron vanishes in the heavy 
nucleon limit.   

In the numerical analysis we use the variational {\it Gaussian ansatz} 
\cite{PCQM,Duck} for the quark ground state wave function with the following  
analytical form: 
\begin{eqnarray}\label{Gaussian_Ansatz} 
u_0(\vec{x}) \, = \, N \, \exp\biggl[-\frac{\vec{x}^2}{2R^2}\biggr] \, 
\left(
\begin{array}{c}
1\\
i \rho \, \vec{\sigma}\vec{x}/R\\
\end{array}
\right) 
\, \chi_s \, \chi_f \, \chi_c 
\end{eqnarray}      
where $N=[\pi^{3/2} R^3 (1+3\rho^2/2)]^{-1/2}$ is a constant fixed by the 
normalization condition $\int d^3x \, u^\dagger_0(x) \, u_0(x) \equiv 1$; 
$\chi_s$, $\chi_f$, $\chi_c$ are the spin, flavor and color quark wave 
functions, respectively. Our Gaussian ansatz contains two model parameters: 
the dimensional parameter $R$ and the dimensionless parameter $\rho$. 
The parameter $\rho$ can be related to the axial coupling constant $g_A$ 
calculated in the leading-order (or the three quark-core) approximation: 
\begin{eqnarray}
g_A=\frac{5}{3} \biggl(1 - \frac{2\rho^2} {1+\frac{3}{2} \rho^2} \biggr) . 
\end{eqnarray}
Solving this equation in $\rho$, one gets:
\begin{eqnarray}\label{rho_ga}
\rho^2 = 
2 \biggl(\frac{1 - \frac{3}{5}g_A}{1 + \frac{9}{5}g_A} \biggr). 
\end{eqnarray} 
The parameter $R$ can be physically understood as the mean radius of the 
three-quark core and  is related to the charge radius 
$\la r^2_E \ra^P_{LO}$ of the proton in the leading-order (LO) 
approximation as 
\begin{eqnarray}
\la r^2_E \ra^P_{LO} = \int d^3 x \, u^\dagger_0 (\vec{x}) \,\vec{x}^2 \, 
u_0(\vec{x}) \, = \, \frac{3R^2}{2} \, \frac{1 \, + \, \frac{5}{2} \, \rho^2}
{1 \, + \, \frac{3}{2} \, \rho^2}. 
\end{eqnarray}
In our calculations we use the value $g_A$=1.25 obtained in ChPT 
\cite{Gasser1}. Therefore, we have only one free parameter, that is $R$. 
In the numerical studies \cite{PCQM} R is varied in the region from 0.55 fm to 
0.65 fm, which corresponds to a change of $<~r^2_E>^P_{3q-core}$ from 0.5 to 
0.7 fm$^2$. The exact Gaussian ansatz (\ref{Gaussian_Ansatz}) restricts the 
potentials $S(r)$ and $V(r)$ to a form proportional to $r^2$. They are 
expressed in terms of the parameters $R$ and $\rho$ (for details see 
Ref. \cite{PCQM}).  

Using the Gaussian ansatz (\ref{Gaussian_Ansatz}) the proton charge and 
magnetic form factors at zeroth order are determined as \cite{PCQM}: 
\begin{eqnarray}\label{EM_FF}
G_E(-\vec{q}^{\, 2}) &=& \exp\biggl(-\frac{\vec{q}^{\, 2} R^2}{4}\biggr) 
\biggl[ 1 - \frac{\vec{q}^{\, 2} R^2}{4} \kappa \biggr]  , \\
G_M(-\vec{q}^{\, 2}) &=& 
\exp\biggl(-\frac{\vec{q}^{\, 2} R^2}{4}\biggr) 
2m_N R \, \sqrt{\kappa \biggl(1-\frac{3}{2}\kappa\biggr)}, \hspace*{.5cm} 
\kappa = \frac{1}{2} - \frac{3}{10}g_A, \nonumber 
\end{eqnarray} 
which yield finally the electromagnetic mass shift 
\begin{eqnarray}\label{Res_Delta_mn}
\Delta m^{em}_p &=& \frac{\alpha}{R\sqrt{2\pi}} 
\biggl[ 1 - \frac{\kappa}{2} + \frac{3}{16} \kappa^2  
- \frac{34}{9} \kappa \biggl(1 - \frac{3}{2} \kappa\biggr) \biggr] ,\\
\Delta m^{em}_n &=& - \frac{\alpha}{R\sqrt{2\pi}} 
\frac{8}{3} \kappa \biggl(1 - \frac{3}{2} \kappa\biggr) , \nonumber 
\end{eqnarray}
where $\alpha=1/137$ is the fine structure coupling. 
For our set of parameters $g_A=1.25$ and $R=0.6 \pm 0.05$ fm we get 
$\Delta m^{em}_p = 0.54 \pm 0.04$ MeV, $\Delta m^{em}_n = - 0.26 \pm 0.02$ MeV 
and $\Delta m^{em}_n - \Delta m^{em}_p = - 0.8 \pm 0.06$ MeV. 
The uncertainties of our results correspond to the variation of the 
parameter $R$. Our predictions are in qualitative agreement with the results 
obtained by Gasser and Leutwyler using the Cottingham 
formula \cite{Gasser_Leutwyler_PR}: 
$\Delta m^{em}_p = 0.63$ MeV, $\Delta m^{em}_n = -0.13$ MeV,   
$\Delta m^{em}_n - \Delta m^{em}_p =  - 0.76$ MeV. 

To compare our prediction for the electromagnetic mass shifts of the nucleons 
with the result of ChPT \cite{Meissner2}, we recall the part of the ChPT 
Lagrangian \cite{Meissner2} which is responsible for radiative corrections 
\begin{eqnarray}\label{Lagrangian_ChPT} 
\hspace*{-.6cm}
{\cal L}_{ChPT}^{e^2} = e^2 \bar N \biggl\{ 
f_1 \biggl(1 - \frac{\vec{\pi}^{\, 2} - (\pi^0)^{\, 2}}{F^2} \biggr) \, 
+  \, \frac{f_2}{2} \biggl(\tau_3 - \frac{\vec{\pi}^{\, 2}\tau_3 - 
\pi^0 \vec{\pi} \vec{\tau}}{2F^2} \biggr)  + f_3 \biggr\}  N .  
\end{eqnarray}
The $O(p^2)$ low-energy constants (LECs) $f_1$, $f_2$ and $f_3$ contain 
the effect of the direct quark-photon interaction. Matching our results for 
the nucleon mass shifts to the predictions of ChPT \cite{Meissner2} with 
\begin{eqnarray}\label{Delta_m_CHPT}
\hspace*{-1cm} 
\Delta m^{em}_p|_{ChPT} = - 4\pi\alpha \biggl(f_1+f_3+\frac{f_2}{2}\biggr), 
\hspace*{.15cm}  
\Delta m^{em}_n|_{ChPT} = - 4\pi\alpha \biggl(f_1+f_3-\frac{f_2}{2}\biggr) 
\end{eqnarray}
we obtain following relations for the coupling constants $f_1$, $f_2$ and 
$f_3$: 
\begin{eqnarray}\label{f2_estimate} 
\hspace*{-.9cm} 
f_2 &=& - \frac{1}{2R (2\pi)^{3/2}} \biggl[ 1 - \frac{29}{18} \kappa 
+ \frac{89}{48} \kappa^2 \biggr] , \\ 
f_1 + f_3 &=& - \frac{1}{4R (2\pi)^{3/2}} \biggl[ 1 - \frac{125}{18} \kappa + 
\frac{473}{48} \kappa^2 \biggr] . \nonumber 
\end{eqnarray}
Our numerical result for $f_2 = - 8.7 \pm 0.7$ MeV is in good agreement with 
the value of $f_2 = - 8.3 \pm 3.3$ MeV \cite{Meissner2,piP-atom} extracted 
from the analysis of the elastic electron scattering cross section using the 
Cottingham formula \cite{Gasser_Leutwyler_PR}. For $f_1 + f_3$ we get 
$-1.5 \pm 0.1$ MeV.  

We now apply our model to the analysis of radiative corrections to the 
$\pi N$ amplitude at the lowest order in $e^2$, and to the determination of 
$f_1$, $f_3$ and $f_1/f_2$ separately. We denote the corresponding matrix 
element associated with the nucleon flavor transition $N_1 \to N_2$ by 
$M_{N_1N_2}^{(e^2); ij}$. In the Coulomb gauge only six diagrams (Fig.3a-3f) 
contribute to the radiative correction to the $\pi N$ amplitude at threshold. 
The contribution of the other diagrams (Fig.3g-3o) vanishes. The contributions 
of the different diagrams of Fig.3 are as follow: 
\begin{eqnarray}\label{piN_Fig3ab} 
M_{N_1N_2}^{(e^2); ij}\bigg|_{a+b} &=& - \frac{e^2}{F^2} \, \cdot \, 
{^N\la}\phi_0| \int d^4x \int d^4y D_{\mu\nu}(x-y) \bar \psi_0(x) \gamma^\mu \\
&\times& ( T^{ij} G_\psi(x,y) Q + Q G_\psi(x,y) T^{ij}) 
\gamma^ \nu \psi_0(y) |\phi_0{\ra^N} \nonumber 
\end{eqnarray}
for Fig.3a and Fig.3b 
where $T^{ij} = 2\delta^{ij} \tau^3 - \delta^{i3}\tau^j - \delta^{j3}\tau^i$, 
\begin{eqnarray}\label{piN_Fig3c} 
M_{N_1N_2}^{(e^2); ij}\bigg|_{c} &=& \frac{ie^2}{F^2} \, \cdot \, 
{^N\la}\phi_0| \int d^4x \int d^4y D_{\mu\nu}(x-y) \bar \psi_0(x) \gamma^\mu \\
&\times& T^{ij} \psi_0(x) \bar \psi_0(y) \gamma^\nu Q \psi_0(y) |\phi_0{\ra^N} 
\nonumber 
\end{eqnarray}
for Figs.3c, 
\begin{eqnarray}\label{piN_Fig3de} 
M_{N_1N_2}^{(e^2); ij}\bigg|_{d+e} &=& - \frac{e^2}{F^2} \, \cdot \, 
{^N\la}\phi_0| \int d^4x \int d^4y D_{\mu\nu}(x-y) \bar \psi_0(x) 
\gamma^\mu\gamma^5 \\
&\times& (\varepsilon^{3ik} \varepsilon^{3jm} + \varepsilon^{3jk} 
\varepsilon^{3im}) \tau^k G_\psi(x,y) \gamma^\nu \gamma^5 \tau^m 
\psi_0(y) |\phi_0{\ra^N}  \nonumber 
\end{eqnarray}
for Fig.3d and Fig.3e, 
\begin{eqnarray}\label{piN_Fig3f} 
M_{N_1N_2}^{(e^2); ij}\bigg|_{f} &=& \frac{ie^2}{F^2} \, \cdot \, 
{^N\la}\phi_0| 
\int d^4x \int d^4y D_{\mu\nu}(x-y) \bar \psi_0(x) \gamma^\mu\gamma^5 \\
&\times& \varepsilon^{3ik} \varepsilon^{3jm}  \tau^k \psi_0(x) \bar\psi_0(y) 
\gamma^\nu \gamma^5 \tau^m \psi_0(y) |\phi_0{\ra^N} \nonumber 
\end{eqnarray}
for Fig.3f. 

Truncating the quark propagator to the ground state mode we obtain the total 
expression for the $O(e^2)$ $\pi N$ scattering amplitude, at threshold:   
\begin{eqnarray}\label{N1N2} 
M_{N_1N_2}^{(e^2); ij} &=& \frac{\alpha}{(4\pi F)^2} 
\int \frac{d^3q}{\vec{q}^{\,\, 2}} \biggl[ C_{N_1N_2}^{E; ij} 
G_E^2(-\vec{q}^{\, 2}) \, - \,  \frac{\vec{q}^{\,\, 2}}{2 m_N^2} \, 
C_{N_1N_2}^{M; ij} G_M^2(-\vec{q}^{\, 2}) \\
&+& \frac{6}{25} \frac{d^2_+(\vec{q}^{\, 2})}{d^2_-(\vec{q}^{\, 2})} 
C_{N_1N_2}^{A; ij} G_A^2(-\vec{q}^{\, 2}) \biggr] . \nonumber  
\end{eqnarray}
where $G_A$ is the axial charge form factor calculated 
at zeroth order \cite{PCQM} 
\begin{eqnarray}
(\chi^\dagger_{N_{s^\prime}} \sigma^3_N \chi_{N_s}) \, 
(\chi^\dagger_{N_{f^\prime}} \tau^3_N \chi_{N_f}) \, 
&G_A(-\vec{q}^{\, 2})& \,\, = {^N\la}\phi_0|\int d^3x \bar u_0(\vec{x}) 
\gamma^3\gamma^5 \tau^3 u_0(\vec{x})e^{i \vec{q} 
\,\vec{x}}|\phi_0{\ra^N} , \\
&G_A(-\vec{q}^{\, 2})& \,\, = g_A 
\exp\biggl(-\frac{\vec{q}^{\, 2} R^2}{4}\biggr)\, d_-(\vec{q}^{\, 2})\nonumber 
\end{eqnarray} 
and
$$d_{\pm}(\vec{q}^{\, 2}) = 1 \pm \frac{\vec{q}^{\, 2} R^2}{4}
\frac{\kappa}{1 - 2\kappa} .$$
Here $\tau^3_N$ is the nucleon isospin matrix. 
Thus, the contribution of the Coulomb photons to the amplitude 
$M_{N_1N_2}^{(e^2); ij}$ is parametrized by the proton charge form factor 
$(G_E)$, transverse photons are related to the proton magnetic $(G_M)$ and 
axial charge $(G_A)$ form factors. 

The factors $C_{N_1N_2}^E$, $C_{N_1N_2}^M$ and $C_{N_1N_2}^A$ are the 
spin-flavor coefficients
\begin{eqnarray}
\hspace*{-.8cm}
C_{N_1N_2}^{E; ij} &=& <N_1| \sum\limits_{m=1}^3 
\biggl( 2\delta^{ij} \{\tau^3 , Q\} - \delta^{i3} \{\tau^j , Q\} 
- \delta^{j3} \{\tau^i , Q\} \biggr)^{(m)} |N_2> \label{CN1N2_E}\\
&+& 2 <N_1| \sum\limits_{m \, \not\! = n}^3 \biggl( 2\delta^{ij} \tau^3 
- \delta^{i3} \tau^j - \delta^{j3} \tau^i \biggr)^{(m)} Q^{(n)} |N_2> , 
\nonumber\\ 
\hspace*{-.8cm}
C_{N_1N_2}^{M; ij} &=& <N_1| \sum\limits_{m=1}^3 \biggl(2\delta^{ij} 
\{\tau^3 , Q\} - \delta^{i3} \{\tau^j , Q\}  
- \delta^{j3} \{\tau^i , Q\} \biggr)^{(m)} |N_2> \label{CN1N2_M}\\
\hspace*{-.8cm}
&+& \frac{2}{3} <N_1| \sum\limits_{m \, \not\! = n}^3 \biggl( 2\delta^{ij} 
\tau^3 - \delta^{i3} \tau^j - \delta^{j3} \tau^i \biggr)^{(m)} 
Q^{(n)} \vec{\sigma}^{(m)} \vec{\sigma}^{(n)} |N_2> , \nonumber\\
\hspace*{-.8cm}
C_{N_1N_2}^{A; ij} &=& 6 (\delta^{ij} - \delta^{i3} \delta^{j3}) 
<N_1| \biggl( \sum\limits_{m=1}^3  {\rm I}^{(m)} 
+ \frac{1}{9} \sum\limits_{m \, \not\! = n}^3 \vec{\tau}^{\, (m)} 
\vec{\tau}^{\, (n)} \vec{\sigma}^{(m)} \vec{\sigma}^{(n)} \biggr) |N_2> ,
\label{CN1N2_A} 
\end{eqnarray}
where the notation $\{\tau, Q\}$ denotes the anticommutator. The matrix 
elements of Eqs. (\ref{CN1N2_E})-(\ref{CN1N2_A}) can be easily evaluated 
by using the $SU(6)$ spin-flavor wave functions of the nucleon. 

Finally the first-order radiative corrections to the $\pi N$ scattering 
amplitude at threshold are given by 
\begin{eqnarray}\label{M_inv_PCQM} 
T_{\pi N}^{(e^2); ij} &=& -\frac{1}{(4\pi)^3} 
\int\frac{d^3 q}{\vec{q}^{\,\, 2}} 
\bigg\{ T^{ij}_{f_1} \, \biggl[ [G_E^p(-\vec{q}^{\, 2})]^2  
- \frac{19\vec{q}^{\,\, 2}}{6m_N^2} [G_M^p(-\vec{q}^{\, 2})]^2 
+ \frac{114}{25} \frac{d^2_+(\vec{q}^{\, 2})}{d^2_-(\vec{q}^{\, 2})}  
G_A^2(-\vec{q}^{\, 2}) \biggr]\\ 
&+& T^{ij}_{f_2} \, \biggl[ [G_E^p(-\vec{q}^{\, 2})]^2 
- \frac{5\vec{q}^{\,\, 2}}{18m_N^2} 
[G_M^p(-\vec{q}^{\, 2})]^2 \biggr] \biggr\} \nonumber\\
&=& - \frac{1}{8R} \frac{1}{(2\pi)^{3/2}} \biggl\{ T^{ij}_{f_1} 
\biggl[ \frac{41}{3} - \frac{115}{2}\kappa + \frac{953}{16} \kappa^2 \biggr] 
+ T^{ij}_{f_2} \, \biggl[ 1 - \frac{29}{18} \kappa + 
\frac{89}{48} \kappa^2  \biggr] \biggr\} \nonumber . 
\end{eqnarray}
where 
\begin{eqnarray}\label{R_f1_f2}
\hspace*{-.9cm} 
T^{ij}_{f_1} = - \frac{8\pi\alpha}{F^2}  
\, ( \, \delta^{ij} - \delta^{i3} \delta^{j3} \, ) \,\,\,\,\, \mbox{and} 
\,\,\,\,\, T^{ij}_{f_2} = - \frac{4\pi\alpha}{F^2} 
\, (\, 2 \,\delta^{ij} \tau^3 - \delta^{i3} \tau^j - \delta^{j3} \tau^i \, ) . 
\end{eqnarray}
Again, as in the case of electromagnetic mass shifts, the amplitude 
$M_{inv}^{e^2 \pi N}$ is gauge-independent. 
In ChPT the corresponding amplitude is given by \cite{Meissner2} 
\begin{eqnarray}\label{M_inv_ChPT}
\hspace*{-.4cm} 
T_{\pi N}^{(e^2); ij}\large|_{ChPT} \, = \, f_1 \, T^{ij}_{f_1} \, 
+ \, \frac{f_2}{4} \, T^{ij}_{f_2} . 
\end{eqnarray}
Comparing Eqs. (\ref{M_inv_PCQM}) and (\ref{M_inv_ChPT}) we get the same 
expression for $f_2$ as already obtained from the electromagnetic 
mass shift (\ref{f2_estimate}). We also deduce the following relations: 
\begin{eqnarray}\label{f_1andf_2} 
f_1 &=& - \frac{1}{8R (2\pi)^{3/2}} 
\biggl[ \frac{41}{3} - \frac{115}{2} 
\kappa + \frac{953}{16} \kappa^2 \biggr] , \nonumber \\ 
f_3 &=& \frac{1}{8R (2\pi)^{3/2}} \biggl[ \frac{35}{3} - \frac{785}{18} \kappa 
+ \frac{1913}{48} \kappa^2 \biggr] , \\ 
\frac{f_1}{f_2} &=& \frac{\displaystyle{\frac{41}{12} - \frac{115}{8} 
\kappa + \frac{953}{64} \kappa^2}}{\displaystyle{1 - \frac{29}{18} \kappa 
+ \frac{89}{48} \kappa^2 }} . \nonumber 
\end{eqnarray} 
The predicted ratio for $f_1/f_2$ depends on only one model parameter $\rho$ 
(or $\kappa$) which is related to the axial nucleon charge $g_A$ calculated 
at zeroth order (see Eq. (\ref{rho_ga})). In addition, the constants 
$f_1$, $f_2$ and $f_3$ depend on the size parameter $R$ of the bound quark. 
For our "canonical" set of parameters, $g_A=1.25$ and $R=0.6 \pm 0.05$ fm, 
used in the calculations of nucleon electromagnetic form factors and 
meson-baryon sigma terms \cite{PCQM}  we obtain: 
\begin{eqnarray}
& &f_1 = -19.5 \pm 1.6 \,\,\, \mbox{MeV}, \hspace*{1cm}
   f_2 = -8.7 \pm 0.7 \,\,\, \mbox{MeV}, \\
& &f_3 = 18 \pm 1.5 \,\,\, \mbox{MeV}, \hspace*{1.6cm} 
\frac{f_1}{f_2} = 2.2 . \nonumber 
\end{eqnarray}
Using these values of $f_1$ and $f_2$ we can estimate the isospin-breaking 
correction to the energy shift of the $\pi^- p$ atom in the $1s$ state. 
The strong energy-level shift $\epsilon_{1s}$ of the $\pi^- p$ atom is given 
by the model-independent formula \cite{piP-atom}: 
\begin{eqnarray}
\epsilon_{1s} = \epsilon_{1s}^{LO} +  \epsilon_{1s}^{NLO} = 
\epsilon_{1s}^{LO} (1+\delta_{\epsilon}) .  
\end{eqnarray} 
where the leading order (LO) or isospin-symmetric contribution is 
$\epsilon_{1s}^{LO}$ and the next-to-leading order (NLO) or isospin-breaking 
contribution is $\epsilon_{1s}^{NLO}$. 
The quantity $\epsilon_{1s}^{LO}$ is expressed with the help of the 
well-known Deser formula \cite{Deser} in terms of the $S$-wave $\pi N$ 
scattering lengths $a_{\frac{1}{2}}$ and $a_{\frac{3}{2}}$ with 
\begin{eqnarray}
\epsilon_{1s}^{LO} = - 2\alpha^3\mu_c^2 {\cal A}_{str} \hspace*{1cm} 
\mbox{and} \hspace*{1cm} 
{\cal A}_{str} = \frac{1}{3} \, ( 2 a_{\frac{1}{2}} + a_{\frac{3}{2}} ) .
\end{eqnarray}
The reduced mass of the $\pi^- p$ atom is denoted by 
$\mu_c=m_p M_{\pi^+}/(m_p + M_{\pi^+})$ and 
${\cal A}_{str}=(88.4\pm 1.9)\times 10^{-3} M_{\pi^+}^{-1}$ 
is the strong (isospin-invariant) regular part of the $\pi^- p$ scattering 
amplitude at threshold \cite{PSI} (for the definitions of these quantities see 
Ref. \cite{piP-atom}). In ChPT the quantity $\delta_{\epsilon}$, the ratio 
of NLO to LO corrections, is expressed in terms of the LECs $c_1$,  
$f_1$ and $f_2$ \cite{piP-atom}:  
\begin{eqnarray}\label{delta_epsilon}
\delta_{\epsilon} &=& \frac{\mu_c}{8\pi M_{\pi^+} F^2_\pi {\cal A}_{str}} 
[ 8c_1 (M_{\pi^+}^2 - M_{\pi^0}^2) - e^2 (4 f_1 + f_2) ] 
- 2\alpha\mu_c({\rm ln}\alpha - 1) {\cal A}_{str} \, . 
\end{eqnarray}
The quantity $c_1$ is the strong LEC from the ChPT Lagrangian 
\cite{Meissner1,Leutwyler1} and $F_\pi=92.4~{\rm MeV}$ is the physical 
value of the pion decay constant \cite{piP-atom}. In Ref. \cite{PCQM} we 
obtained $c_1 = - 1.2 \pm 0.1$ GeV$^{-1}$ using the PCQM approach. 
Our prediction for $c_1$ is close to the value $c_1 = - 0.9 m_N^{-1}$ deduced 
from the $\pi N$ partial wave analysis KA84 using baryon chiral perturbation 
theory \cite{Leutwyler1}. Substituting the central values for our couplings 
$f_1 = -19.5$ MeV, $f_2 = - 8.7$ MeV and $c_1 = -1.2 $ GeV$^{-1}$ into 
Eq. (\ref{delta_epsilon}), we get $\delta_{\epsilon} = - 2.8 \cdot 10^{-2}$.  
Our estimate is comparable to a prediction based on a potential model for the 
$\pi N$ interaction \cite{PSI}: $\delta_{\epsilon} = -2.1 \cdot 10^{-2}$. 
 
\section{Summary and conclusion}

Our results can be summarized as follows. Starting from the perturbative 
chiral quark model (PCQM), which, by construction, includes confinement and 
in general a nonlinear pion-quark interaction respecting global chiral 
symmetry, we made a perturbative expansion of the Lagrangian with respect to 
$1/F$. In the second order $O(1/F^2)$ one gets an interaction as well as 
a contact term quadratic in the pion field. We then show that the PCQM, both 
on the Lagrangian level and for observables, is fully equivalent to a 
Weinberg type model with an axial-vector pion-quark coupling and the explicit 
Weinberg-Tomozawa (WT) term. This connection is based either on an unitary 
transformation of the quark fields, which leaves the unperturbed part of the 
Lagrangian unaltered, or by the explicit evaluation of observables, such as 
the low-energy $\pi N$ $S$-wave scattering amplitude. 

As an intermediate result we note that:

i) the PCQM has two Lagrangian manifestations, where, as a matter of 
calculational convenience, either version can be used in the perturbative 
evaluation. 

ii) the full equivalence between two model versions is only given, when 
the full quark propagator is introduced; hence a truncation of the 
intermediate-state sums will lead in general to different results. 

iii) the connection between pseudoscalar and axial vector pion-quark coupling 
is made explicit; in the context of the PCQM this relation is established 
since the quarks remain on their energy-shell and quadratic terms in the pion 
field are kept consistently. 

iv) the PCQM fulfils the low-energy constraints set by the WT result for 
$\pi N$ scattering lengths. 

In a second step we extended the PCQM to include the electromagnetic field. 
Again, both Lagrangian versions are fully equivalent when introducing the 
photon field by minimal substitution. We apply the formalism to radiative 
corrections of low-energy nucleon observables. In the limit of a heavy 
nucleon mass we express the electromagnetic mass shift of nucleons and the 
$e^2$ corrections to the $\pi N$ amplitude in terms of the proton charge form 
factor only. Final expressions can be shown to be independent of the gauge 
choice. The result for the mass shift corrections is consistent with the 
model-independent one obtained in Ref. \cite{Gasser_Leutwyler_PR}. We can 
determine the complete set of the $O(p^2)$ electromagnetic low-energy 
couplings (LECs) $f_1$, $f_2$ and $f_3$ of the chiral perturbation theory 
(ChPT) effective Lagrangian. The magnitude of $f_2$ and its relation to $f_1$ 
and $f_3$ are obtained from an analysis of the nucleon electromagnetic mass 
shift and the leading radiative corrections to the $\pi N$ scattering 
amplitude at threshold. Using our values for $f_1$ and $f_2$ we also predict 
the isospin-breaking correction to the strong energy shift of the $\pi^- p $ 
atom in the $1s$ state. This prediction is extremely important for the 
ongoing experiment "Pionic Hydrogen" at PSI, which plans to measure the 
ground-state shift and width of pionic hydrogen ($\pi^- p$-atom) at the 
$1\%$ level \cite{Gotta}.  

\vspace*{.3cm}
{\it Acknowledgements}. 
We thank A.~Rusetsky for useful discussions. This work was supported by 
the DFG (grant FA67/25-1) and by the DAAD-PROCOPE project.

\newpage 

\vspace*{2cm}

\centerline{FIGURES} 

\vspace*{2cm}

\noindent FIG.1: Diagrams contributing to the $\pi N$ scattering 
amplitude: $s$-channel pole (1a), $u$-channel pole (1b) and 
seagull diagram (1c).  

\vspace*{1cm}

\noindent FIG.2: Diagrams contributing to the electromagnetic mass shift  
of the nucleon: 
\noindent one-body (2a) and two-body diagram (2b). 

\vspace*{1cm}
\noindent FIG.3: Diagrams contributing to the leading $e^2/F^2$ radiative 
corrections to the $\pi N$ amplitude at threshold.

\newpage
\begin{figure}
\vspace*{16cm}
\includegraphics{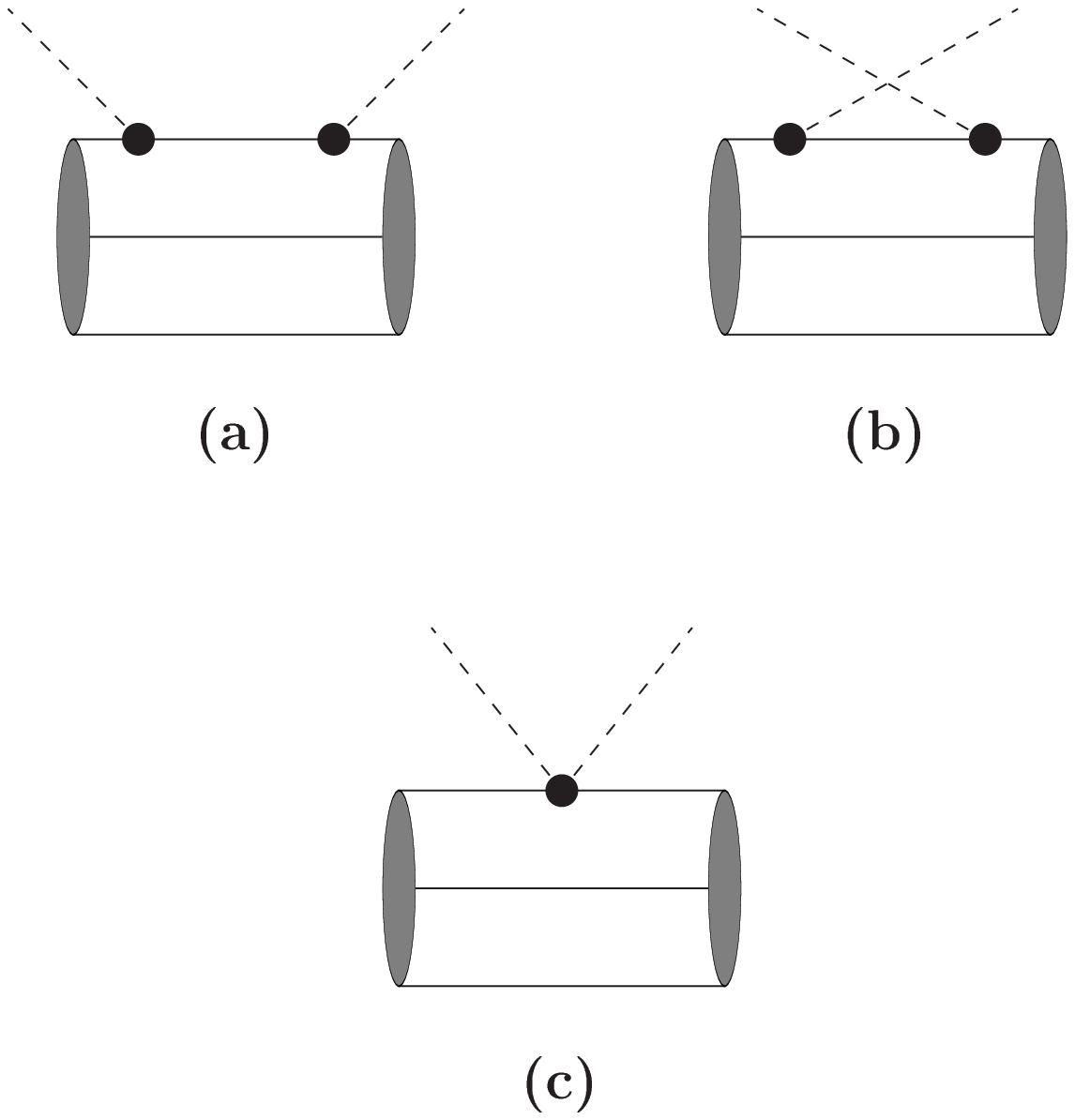}

\vspace*{-4cm}
\hspace*{6cm}{\bf Fig.1}
\end{figure}

\begin{figure}
\vspace*{14cm}
\includegraphics{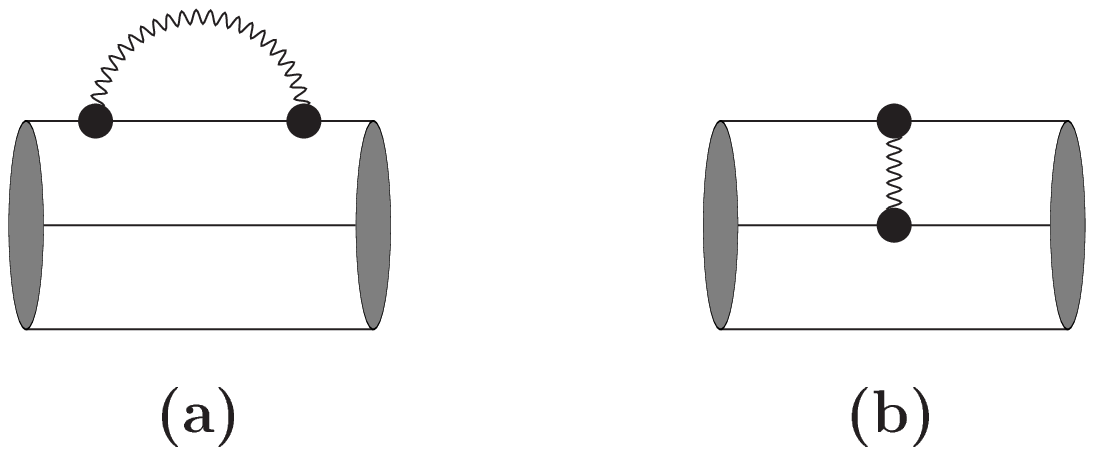}
\vspace*{-7cm}
\hspace*{6cm}{\bf Fig.2}
\end{figure}

\newpage
\begin{figure}
\vspace*{16cm}
\includegraphics{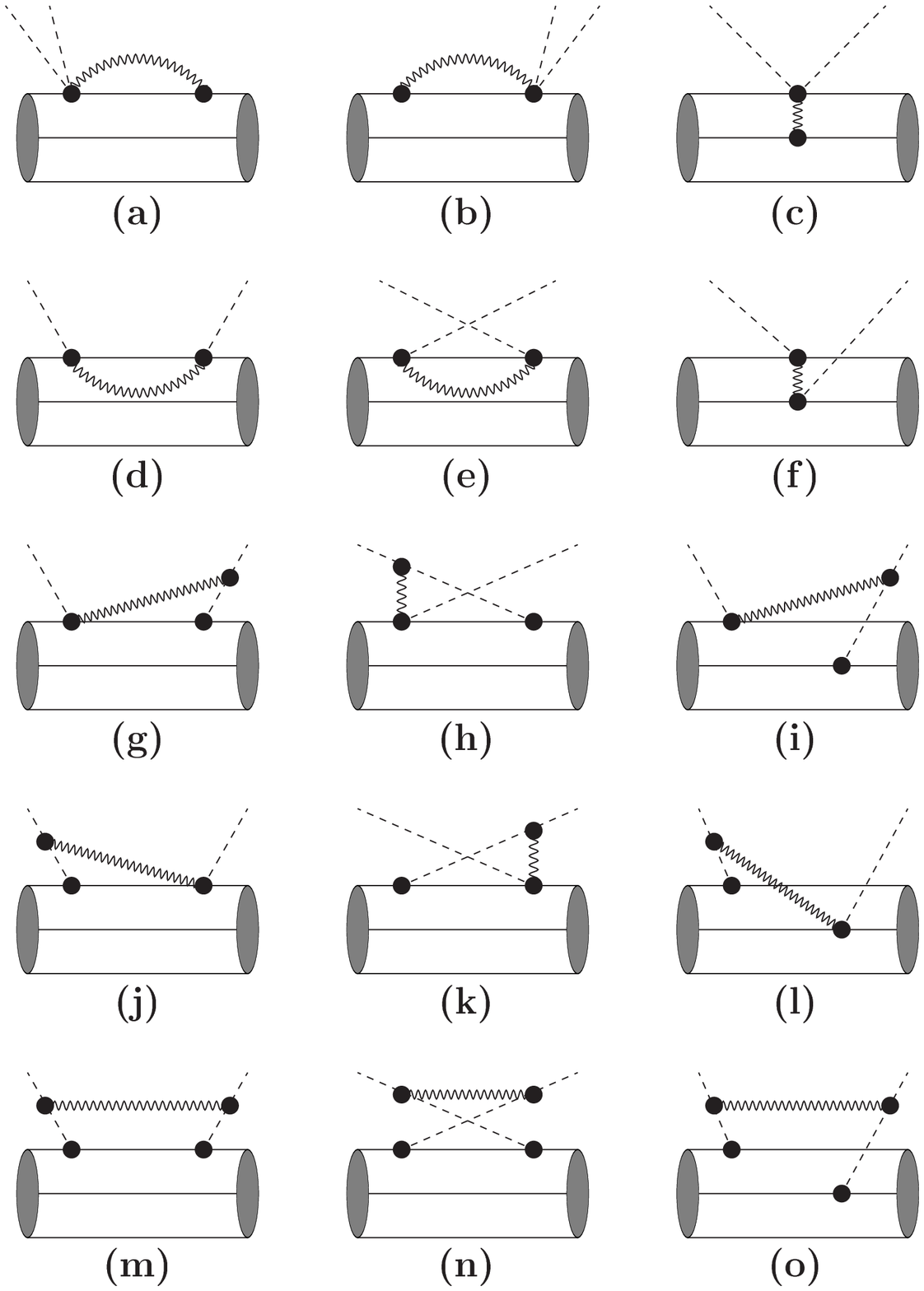}
\vspace*{3cm}
\hspace*{6cm}{\bf Fig.3}
\end{figure}
\end{document}